\documentstyle{mn}

\title{A search for symbiotic behaviour amongst OH/IR colour mimics}

\author[E. R. Seaquist \& R. J. Ivison]
       {E. R. Seaquist and R. J. Ivison\\ Department of Astronomy,
University of Toronto, 60 St.\ George Street, Toronto, Ontario M5S
1A7, Canada}
\date{Accepted ...
      Received ...
      in original form ...}
\pagerange{000--000}

\begin{document}

\maketitle

 \begin{abstract} Recent maser surveys have shown that many potential
OH/IR stars have no OH masers in their circumstellar envelopes,
despite the modest requirements which should be implicitly met by {\it
IRAS} colour-selected candidates. It has been suggested that these
OH/IR colour mimics must have a degenerate companion which dissociates
OH molecules and disrupts the masing action, ie.\ that they are
related to symbiotic Miras. Coincidentally, there is a paucity of
long-period symbiotic Miras and symbiotic OH/IR
stars. Phenomonologically, those that are known seem to cluster in the
zone where field Miras transform into OH/IR stars. If it could be
proven that OH/IR colour mimics contain a degenerate star, that
observable evidence of this star is hidden from view by circumstellar
dust whilst it slowly accretes from the wind of its Mira companion,
then we have an excellent explanation for not only the existence of
OH/IR colour mimics, but also for the low observed frequency of
symbiotic OH/IR stars and the common occurrence of very slow novae in
long-period symbiotic Miras. Here, we employ radio continuum radiation
(which should escape unhindered from within the dust shells) as a
simple probe of the postulated hot degenerate companions which would
inevitably ionize a region of their surrounding gas. We compare the
radio and infrared properties of the colour mimics with those of
normal symbiotic Miras, using the strong correlation between radio and
mid-IR emission in symbiotic stars. We show that if a hot companion
exists then, unlike their symbiotic counterparts, they must produce
radiation-bounded nebulae.  Our observations provide no support for
the above scenario for the lack of observed masers, but neither do
they permit a rejection of this scenario.
 \end{abstract}

\section{Introduction}

As a Mira evolves from short to longer pulsational periods, its
mass-loss rate increases dramatically (Schild 1989). The object
evolves from a classical Mira into a variable OH/IR source, a label
which denotes the simultaneous presence of copious far-infrared
emission and a 1,612-MHz OH maser. Circumstellar (CS) dust, formed in
the wind of these pulsating giants, protects their molecules against
photodissociation by interstellar UV. Its associated infrared (IR)
emission also pumps the OH molecules for their stimulated emission.
Stars {\it without} dust have no masers -- there is no defensive
shield. We therefore expect 1,612-MHz masers to be associated with
strong, stellar far-IR sources and vica versa (particularly those that
are oxygen-rich as a surplus of oxygen is required for the OH masers).

Potential OH/IR stars can be identified in the Infrared Astronomical
Satellite ({\it IRAS}) Point Source Catalog ({\it PSC}) by searching
for the signature of a dust shell -- ie.\ its colour (eg.\ Engels et
al.\ 1984; Gaylard et al.\ 1989). This approach has added over 1,000
new examples of the OH/IR phenomenon to the literature by successfully
finding their 1,612\,MHz OH masers (Eder, Lewis \& Terzian 1988; Lewis,
Eder \& Terzian 1990; Lintel-Hekkert et al.\ 1991).

Despite the apparent success of colour selection techniques, recent
maser surveys (eg.\ Lewis 1992a) have demonstrated that many potential
OH/IR stars have no 1,612-MHz OH masers despite the modest
requirements of a flux of 35- and 53-micron pump photons and a suitable
column density of OH molecules (which should be implicitly met by {\it
IRAS PSC} colour-selected candidates). Many also lack mainline (1,665
and 1,667\,MHz) OH and 22-GHz water masers. For these objects the
absence of a detectable OH maser has been shown to be independent of
telescope sensitivity and it is not attributable to IR variability
(less than 2 percent are detected when observed for the second time).
Furthermore, many of the candidate stars are O-rich (LRS types 2n--3n,
ie.\ the 9.7-micron silicate feature is present in either emission or
absorption).  Lewis (1992a) and Lewis \& Engels (1993) demonstrated
that 17\% of LRS type 2n--3n sources with colours suggestive of CS
dust shells have no OH or H$_2$O masers.

It has been suggested (Lewis 1992b) that the natural explanation for
the ``OH/IR colour mimics'' is that they are systems with a degenerate
companion -- a {\it local} source of UV which disrupts the CS shell
and prevents the masing action, ie.\ that the colour mimics are
closely related to symbiotic Miras.

According to the conventional picture of the symbiotic Miras (which
are probably represented exclusively by the D(usty)-type sub-class of
symbiotic stars), a hot, degenerate star ionizes the Mira wind giving
rise to a forest of emission lines in the optical/UV (together with
the normal far-IR dust emission and semi-regular pulsations expected
from the cooler star). The hot companion exhibits unusually slow
nova-like outbursts (eg.\ RR Tel), possibly the consequence of shell
flashes resulting from accretion from the Mira wind (see eg.\ Allen \&
Wright 1988). The symbiotic Miras are less numerous than the
S(tellar)-type sub-class which contain a first-ascent red giant whose
stellar atmosphere is the dominant contributor to the observed IR
emission.

The known symbiotic Miras appear to be generally devoid of maser
emission, according to a number of recent searches, and the hypothesis
is that the hot companion is responsible for either the dissociation
of the relevant molecules or for disruption of the maser emission
mechanism (eg.\ Norris et al.\ 1984). In support of this picture,
Lewis, Hajian \& Terzian (1992) searched the International Ultraviolet
Explorer (IUE) data archive for a small, colour-selected sample of
stars without OH or H$_2$O masers (many, in fact, were bona fide
symbiotic stars) and found that most exhibited a strong UV
continuum. Thus it is reasonable to suspect that the absence of maser
emission may be associated with the presence of a hot companion.

Phenomenologically, symbiotic Miras cluster in the zone where field
Miras transform into OH/IR stars (Schild 1989).  The correspondingly
high mass-loss rate of D-type symbiotic Miras is generally thought to
lead to stronger optical/UV line emission than is seen from the
S-types. This effect enhances a symbiotic Mira's chance of discovery
in objective prism surveys. However, it is possible that these
symbiotics spend many years prior to outburst in hibernation
(resembling solitary Miras in most respects) as the hidden companion
accretes slowly from the Mira's wind. In this case the hot companion
and associated emission lines would be totally obscured optically by
the dusty CS shell of the Mira. Schild (1989) has suggested that the
obscuration of the hot component by the dusty CS shell produces a
selection against the detection of symbiotic binaries with Mira or
OH/IR stars cool components and may therefore be responsible for their
paucity. The suggestion by Lewis (1992b) is that the hot companion may
never-the-less be effective in destroying any maser action, even though
it is not optically visible. Recently, Lewis et al.\ (in preparation)
obtained new IUE data for an unbiased sample of colour mimics (our
sample -- see later) they failed to detect UV continuum emission,
presumably because of obscuration by CS dust. Lewis suggests with
resignation that `D-type symbiotic stars can be identified among
sources with thick, opaque dust shells by a persistent absence of
appropriate masers.'

Thus these colour-mimics have implications for a proper understanding
of the evolution of OH/IR stars and for the behaviour of symbiotic
binaries. To properly test the hypothesis of Lewis (and thereby
address some of the questions posed above) it is necessary to somehow
probe through the CS dust shells of OH/IR colour mimics. Here, we
describe a search for free-free continuum at cm wavelengths from gas
ionized by the hot, stellar component. Radio continuum radiation
should easily escape from within a CS dust shell.

Observations of radio continuum emission have long been exploited to
acquire knowledge of of processes in symbiotics (eg.\ Seaquist \&
Taylor 1990) and it is worth noting that all of the symbiotic Miras
accessible from the VLA have been detected at 3.6\,cm. Our method of
determining whether the degenerate companions truly exist is therefore
to compare the continuum flux densities of the OH/IR colour mimics
with continuum measurements of symbiotic stars from the northern- and
southern-sky surveys by Seaquist, Krogulec and Taylor (1993, hereafter
SKT93) and Ivison \& Seaquist (in preparation).

In what follows we describe our continuum measurements of 15 {\it IRAS
PSC} OH/IR colour mimics (massive, oxygen-rich, LRS types 2n--3n)
using the NRAO\footnote{The National Radio Astronomy Observatory
(NRAO) is operated by Associated Universities Inc., under a
cooperative agreement with the National Science Foundation.} Very
Large Array (VLA).

\section{Observations and Results}

 \begin{table*}
 \caption{The sample of OH/IR colour mimics and results of the VLA
observations}
 \begin{tabular}{lcccc}
\hline
          &    &          &         &                    \\
Source name(s)&{\it IRAS}&\multicolumn{2}{c}{Coordinates (B1950)}&Flux Density
at\\
          &LRS &R.A.      &dec      &3.6\,cm/8.44\,GHz  \\
          &type&h m s     &$^{\circ} \; ' \; ''$&($\mu$Jy)\\
          &    &          &                             \\
\hline
          &    &          &                             \\
06181+0406&23  &06 18 07.3&+04 06 36&$3\sigma < 59$     \\
06582+1507&22  &06 58 17.3&+15 07 58&$3\sigma < 81$     \\
18281+2149 (AC Her; BD$+$21 3459)&25  &18 28 08.8&+21 49 50&$3\sigma < 67$\\
18586+0106&35  &18 58 38.1&+01 06 57&$3\sigma < 229$\\
19135+0931 (AFGL 2350)&24  &19 13 30.6&+09 31 32&$3\sigma < 56$\\
19282+2253&26  &19 28 12.1&+22 53 39&$3\sigma < 60$     \\
19310+1745&39  &19 31 02.8&+17 45 31&$3\sigma < 73$     \\
19420+3318&24  &19 42 01.5&+33 18 08&$3\sigma < 67$     \\
19548+3035 (AFGL 2477)&21  &19 54 48.8&+30 35 53&$3\sigma < 61$\\
19558+3333&31  &19 55 53.3&+33 33 11&$440\pm90^{\star}$ \\
19573+3143&29  &19 57 18.0&+31 43 55&$3\sigma < 69$     \\
19584+2652&32  &19 58 26.7&+26 52 18&$3\sigma < 55$     \\
19586+3637 (V1511 Cyg; IRC$+$40371)&29  &19 58 39.1&+36 37 50&$3\sigma < 46$\\
20217+3330&24  &20 21 42.7&+33 30 53&$3\sigma < 69$     \\
20220+3404&34  &20 22 04.2&+34 04 53&$3\sigma < 69$     \\
          &    &          &                             \\
\hline
 \end{tabular}

\noindent
$^{\star}$\,R.A.\ $= 19^{\rm h}\, 58^{\rm m}\, 39.4^{\rm s}$;
dec $= 36^{\circ}\, 37'\, 50''$ (B1950)
 \end{table*}

 \begin{figure}
 \vspace{14cm}
 \caption{{\it IRAS} two-colour diagrams of (a) the 15 OH/IR colour mimics
observed by the VLA and (b) the 53 symbiotic stars detected in two or
more of the 12-, 25- or 60-micron {\it IRAS} bands. Both sets of data are
superimposed upon the colour distribution of normal OH/IR stars
(Lewis \& Engels 1993). Four-pointed stars: OH/IR colour mimics;
filled circles: D-type symbiotics; empty circles: S-type symbiotics;
filled diamonds: D$'$-type symbiotics.}
 \end{figure}

Observations of 15 OH/IR colour mimics at 3.6\,cm were carried out
during 1993 September 03--04 using the VLA in a hybrid of the C and D
configurations with the northern arm of the array longer than the
eastern and western arms. The total bandwidth for our observations was
100\,MHz, centred at 8.44\,GHz (3.6\,cm). Two separate IF pairs were
employed, each containing right and left circular polarizations, thus
four measurements were recorded at 15-s intervals for each antenna.
Later, during mapping, the four measurements were averaged for each
time interval. The FWHM of the synthesized beam (averaged between
major and minor axes) was about 7 arcsec.

The observing procedure was standard in most respects. Observations
(four 10-min snapshots centred on a position 10 arcsec north of each
target star) were sandwiched and interspersed with brief measurements
of bright, unresolved calibrators. The flux densities of the targets
were tied to those of their nearby calibrators which, in turn, were
tied to the flux density of 3C\,286 (5.27\,Jy at 3.6\,cm).

Calibration closely followed the recipes laid down in the NRAO AIPS
Cookbook. The phase and amplitude solutions were extremely stable
throughout the 15-hr observing period. Maps of each target were then
made using the MX routine within AIPS. The maps had dimensions $256
\times 256$ pixels, usually with 1.00 or 2.25 arcsec$^2$ pixels, and
we employed up to 5,000 CLEAN iterations. The resulting noise level
was very close to the theoretical limit, usually between 15 and
25\,$\mu$Jy.

Larger maps were made for targets with bright, confusing objects
nearby. For example, a planetary nebula (He $1-4$, PK $68+1^{\circ}2$)
was visible $\sim$3\,arcmin to the NNE of the colour mimic 19573+3143
(at R.A.\ (B1950) $= 19^{\rm h}\, 57^{\rm m}\, 20.5^{\rm s}$; dec $=
31^{\circ}\, 46'\, 23''$) and so a $17' \times 17'$ map was created.

Our search list began as the sample of O-rich colour mimics ({\it
IRAS} {\it PSC} LRS types 2n--3n) listed in Table 7 of Lewis
(1992a). Lewis \& Engels (1993) later pruned the list from 26 to 15
objects by obtaining very high sensitivity OH mainline and H$_2$O
22-GHz line data with the Arecibo and Effelsberg antennas,
respectively. The list of remaining targets (together with {\it IRAS}
coordinates and alternative names) is given in Table 1 together with
the 3.6-cm flux density for the one detected and upper limits for the
remaining 14 objects. We accepted a detection if the radio position
was within the 95 percent significance level error ellipse of the {\it
IRAS} coordinates. The upper limits are three times the rms noise on
the map.

\section{Discussion}

\subsection{{\it IRAS} colours}

Figure 1(a) shows an {\it IRAS} two-colour diagram of the OH/IR colour
mimics in our sample, superimposed upon the colour distribution of
typical OH/IR stars. The data were taken directly from the {\it IRAS
PSC}.  For comparison, the colour distribution of 53 symbiotics (those
detected in two or more of the 12-, 25- and 60-micron bands) is shown
in Figure 1(b). Of the $\sim165$ known symbiotic stars, 68 were
detected in at least one band by {\it IRAS} (Munari \& Ivison, in
preparation).

The $(25-12)$ and $(60-25)$ colours are defined in the $\nu S_{\nu}$
formalism, for an assumed blackbody temperature of 300\,K, as
 \begin{equation}
(25-12) = {\rm log}_{10} \left( \frac{S_{25} \times 12 \times 0.89}
{S_{12} \times 25 \times 1.09} \right)
 \end{equation}
 \noindent
and
 \begin{equation}
(60-25) = {\rm log}_{10} \left( \frac{S_{60} \times 25 \times 0.82}
{S_{25} \times 60 \times 0.89} \right)
 \end{equation}
 \noindent
where $S_{12}$, $S_{25}$ and $S_{60}$ are the {\it IRAS} {\it PSC} flux
densities at 12, 25 and 60\,microns.

It is clear from Figure 1 that, in terms of their {\it IRAS} colours,
the OH/IR colour mimics are more closely related to the (D-type)
symbiotic Miras than to the S-type symbiotics. We also note a strong
similarity between the colours of the `yellow' (D$'$-type) symbiotic
stars, specifically V741 Persei, Wray 157 and AS 201, and those of
normal OH/IR stars (cf.\ Kenyon, Fern\'{a}ndez-Castro \& Stencel
1988).

\subsection{OH and H$_2$O masers in symbiotic Miras}

The recent maser-line survey of symbiotic Miras by Seaquist \& Ivison
(in preparation) resulted in the detection of 1,612-MHz OH and
22-GHz H$_2$O masers in H1$-$36 and R~Aqr (see also Ivison, Seaquist
\& Hall 1994). The level of maser emission from H1$-$36 ($\sim250$\,mJy)
would certainly have ensured its detection had it been a target in the
search for OH and H$_2$O emission from OH/IR colour mimics by Lewis \&
Engels (1993).  The detection of H1$-$36 at 1,612\,MHz is ironic because
the work presented by Lewis (1992a) was inspired by the association of
H1$-$36 with {\it IRAS} source 17463$-$3700 and its supposed lack of a
detectable OH maser. R~Aqr is the nearest known symbiotic star ($d =
200$\,pc), and detection of its masers would not have been
possible at the distance of H1$-$36 ($d > 5$\,kpc).

Both R~Aqr and H1$-$36 are sources of intense radio continuum and
optical line emission. It is generally thought that $10^4$--$10^5$\,K
degenerate stars orbit the Mira components, probably with binary
separations in the range 10--1,000\,AU. In the case of H1$-$36, it is
possible that the hot companion lies {\it outside} the dusty CS
envelope as the reddening towards the Mira exceeds that towards the
ionized nebula by $\sim18$\,mag (Allen 1983).

Thus recent maser-line observations have unequivocally shown that dust
can shield some CS hydroxyl and water molecules from dissociation,
even in systems which possess intense local sources of UV. If
degenerate companions are responsible for the lack of masers in OH/IR
colour mimics then it is probably reasonable to assume that the UV
luminosity of these hot stars cannot be much less than that of the
degenerate companion in H1$-$36.

\subsection{Radio continuum studies of symbiotic stars and the colour mimics}

 \begin{figure}
 \vspace{9cm}
 \caption{The continuum flux density at 3.6\,cm versus that at
12\,$\mu$m for the symbiotic stars observed by SKT93 and Ivison
\& Seaquist (in preparation) and the OH/IR colour mimics.
Stars referred to explicitly in the text are labelled.  Four-pointed
stars: OH/IR colour mimics; filled circles: D-type symbiotics; empty
circles: S-type symbiotics; filled diamonds: D$'$-type symbiotics.}
\end{figure}

 \begin{figure}
 \vspace{9cm}
 \caption{The continuum flux density at 3.6\,cm versus that at
25\,$\mu$m for symbiotic stars and OH/IR colour mimics.
Four-pointed stars: OH/IR colour mimics; filled circles: D-type
symbiotics; empty circles: S-type symbiotics; filled diamonds:
D$'$-type symbiotics.}
 \end{figure}

Radio continuum observations of $\sim75$ percent of the known symbiotic
stars have been obtained during recent years by SKT93 and Ivison \&
Seaquist (in preparation) using the VLA and Australia Telescope
Compact Array. Both of these surveys concentrated at a wavelength of
3.6\,cm.

SKT93 investigated the dependence of the radio--IR correlation on IR
waveband, using both flux densities and luminosities for this purpose.
Their work confirmed that a radio--IR correlation exists, and that it
is strongest in the mid-IR (Spearman-Rank correlation coefficients:
0.73 at 12\,$\mu$m; 0.64 at 25\,$\mu$m, both with significance
$>99.9$ percent). This causes no surprise as both the dust and the ionized
gas are thought to form part of the CS envelope in symbiotic binary
systems.

Figures 2 and 3 show the most complete sample of radio and IR
measurements of symbiotic stars to date, a total of 60 sources for
which both {\it IRAS} and 3.6\,cm data exist, together with the
measurements of the colour mimics. {\it IRAS} flux densities for the
symbiotics were drawn from Munari \& Ivison (in preparation) and
Kenyon et al.\ (1988). For the symbiotics, the correlation between
both 12 and 25\,$\mu$m flux densities and the 3.6\,cm flux density is
clear to the eye, especially for the D-types. The correlation improves
further if the very slow novae, V1016 Cyg, HM Sge and RR Tel, are
ignored. We note that the position of H1$-$36 on this diagram offers
some credence to the idea, propounded by Allen (1983), that this
system has also experienced a very slow nova-like event.

It is immediately clear even without recourse to statistical tests
that there is virtually no similarity between our sample of OH/IR
colour mimics and the radio-luminous D-type symbiotic Miras. Nor is
there any similarity between S- or D$'$-type symbiotics and the colour
mimics. This impression is confirmed by a 2-D Kolmogorov-Smirnov test
(ignoring upper limits for the symbiotics whilst treating upper limits
for the colour mimics as detections) which yields a probability $p <
10^{-7}$ that the two samples are drawn from the same population. A
1-D test using survival analysis methods (Feigelson \& Nelson 1985;
Isobe, Feigelson \& Nelson 1986) on a sample restricted to the range
of IR flux densities occupied by the color mimics yields $p <
10^{-4}$. All upper limits are included in the latter test. In only
one case, that of 19558$+$3333, is there any possibility of an ionized
nebula approaching the scale of those found in symbiotic Miras. Even
for 19558$+$3333, the 3.6-cm flux density is more than an order of
magnitude lower than symbiotic Miras with equivalent levels of mid-IR
emission.

Thus it is clear that if the colour mimics possess hot companions,
then the size of the ionized zone is relatively small. The radio
detected symbiotics are nearly all density bounded (SKT93), and it is
therefore likely that the colour mimics are radiation bounded. In
terms of the binary model considered by Taylor \& Seaquist (1984) this
is equivalent to stating that the ionization parameter $X < 1/3$,
where $X$ measures essentially the ratio of the UV luminosity to the
available mass in the envelope. It is possible to estimate the upper
limit on the radio continuum flux density for such a system if the
distance and binary separation are assumed.  A rough estimate of the
distance for the colour mimics may be obtained by comparing them with
the group of D-type symbiotics in Figures 2 and 3 occupying the same
range in IR flux density and a broadly similar range in colour.  The
median distance for symbiotics in this group for which there are
available distance estimates (eg.\ SKT93) is about 2\,kpc. Then for an
assumed binary separation of $5 \times 10^{14}$\,cm, the binary model
yields a 3.6-cm flux density less than 100\,$\mu$Jy, not very
different from the observed limits for the colour mimics.  Note also
that if R~Aqr, a probable radiation-bounded system (SKT93), were
placed at 2\,kpc, then its continuum flux density would be near our
detectable limit, and its H$_2$O maser emission would also be
undetectable, though its SiO maser emission might be detectable.
Furthermore, a line of slope unity through R~Aqr in Figures 2 and 3
passes close to the one detection in our sample, thus if R~Aqr were
displaced in distance so that its IR flux matched that of the detected
object, its radio flux would be comparable to our detected source,
suggesting perhaps that most of the colour mimics are strongly
radiation bounded if they contain a hot component at all. This may be
due either to an underluminous hot companion, excessive mass-loss rate
or small binary separation when compared to the D-type symbiotics of
similar colour.

\section{Conclusions}

The search for 3.6\,cm free-free emission from a sample of OH/IR
colour mimics with no OH masers yielded negative results with possibly
one exception. The conclusion is that for all but one of these objects
the suspected hot companion does not ionize sufficient gas to reveal
its presence as a radio source. A comparison with radio-luminous
D-type symbiotics shows that the colour mimics do not belong to the
D-type sub-class with density-bounded envelopes. It is possible
however that they are radiation-bounded D-type symbiotics similar to R
Aqr, whose continuum emission would be only marginally detectable, and
whose H$_2$O maser would be undetectable at the distances of the known
D-type systems.

Thus our results do not exclude the possibility that the colour mimics
possess hot companions provided they produce radiation-bounded nebulae
under these circumstances. Our results therefore provide no support
for the hypothesis that molecular masers are inhibited by UV radiation
from a hot companion. Neither do they contradict the hypothesis since
radiation-bounded nebulae trap only UV emission shortward of the Lyman
continuum. It is noteworthy, however, that a 1,612-MHz OH maser has now
been detected in the distant and radio-luminous symbiotic Mira, H1$-$36,
indicating in this case that UV does not inhibit the maser in this
object. Perhaps the only means to determine whether these OH/IR colour
mimics are ``closet D-type symbiotics'' is to monitor them for the
slow nova-type outbursts which characterize these systems.

\subsection*{Acknowledgments}

The authors would like to acknowledge the hospitality the
millimetre-wave group at the California Institute of Technology where
a large fraction of this work was undertaken. We also acknowledge use
of the survival analysis software package provided by Eric
Feigelson. This work was supported by an operating grant to E.\,R.\
Seaquist from the Natural Sciences and Engineering Research Council of
Canada.

\bsp

\end{document}